\documentclass[sigconf]{acmart}

\AtBeginDocument{%
  }

\renewcommand\footnotetextcopyrightpermission[1]{} 

\thanks{This paper has been accepted at Supercomputing 2025.}
\setcopyright{acmcopyright}
\copyrightyear{2025}
\acmYear{2025}
\setcopyright{cc}
\setcctype{by-nc-sa}
\acmConference[SC '25]{The International Conference for High Performance Computing, Networking, Storage and Analysis}{November 16--21, 2025}{St Louis, MO, USA}
\acmBooktitle{The International Conference for High Performance Computing, Networking, Storage and Analysis (SC '25), November 16--21, 2025, St Louis, MO, USA}
\acmDOI{10.1145/3712285.3759778}
\acmISBN{979-8-4007-1466-5/2025/11}

\keywords{GPUs, SSDs, Asynchronous I/O, Software-managed cache, Memory hierarchy, Storage systems}

\usepackage{booktabs}
\usepackage{natbib}
\usepackage{algorithm}
\usepackage{algpseudocode}

\usepackage{listings}

\lstdefinelanguage{CUDA}{
  morekeywords={
    __global__, __device__, __host__, __shared__, __constant__, define, void, uint32, AgileLockChain,  AGILE_CTRL, AgileBufPtr, AGILE_HOST, SHARE_TABLE_IMPL, GPU_CACHE_IMPL, auto, int, class, public, char,
    __syncthreads, threadIdx, blockIdx, blockDim, gridDim
  },
  sensitive=true,
  morecomment=[l]{//},
  morecomment=[s]{/*}{*/},
  morestring=[b]",
}

\begin{document}

\title{AGILE: Lightweight and Efficient Asynchronous GPU-SSD Integration}


\author{Zhuoping Yang}
\affiliation{%
  \institution{Brown University}
  \city{Providence}
  \state{Rhode Island}
  \country{USA}
}
\email{zhuoping_yang@brown.edu}

\author{Jinming Zhuang}
\affiliation{%
  \institution{Brown University}
  \city{Providence}
  \state{Rhode Island}
  \country{USA}
}
\email{jinming_zhuang@brown.edu}

\author{Xingzhen Chen}
\affiliation{%
  \institution{Brown University}
  \city{Providence}
  \state{Rhode Island}
  \country{USA}
}
\email{xingzhen_chen@brown.edu}

\author{Alex K. Jones}
\affiliation{%
  \institution{Syracuse University}
  \city{Syracuse}
  \state{New York}
  \country{USA}
}
\email{akj@syr.edu}

\author{Peipei Zhou}
\affiliation{%
  \institution{Brown University}
  \city{Providence}
  \state{Rhode Island}
  \country{USA}
}
\email{peipei_zhou@brown.edu}

\renewcommand{\shortauthors}{Zhuoping Yang et al.}


\setlength{\textfloatsep}{5pt}
\begin{abstract}

GPUs are critical for compute-intensive applications, yet emerging workloads such as recommender systems, graph analytics, and data analytics often exceed GPU memory capacity. Existing solutions allow GPUs to use CPU DRAM or SSDs as external memory, and the GPU-centric approach enables GPU threads to directly issue NVMe requests, further avoiding CPU intervention. However, current GPU-centric approaches adopt synchronous I/O, forcing threads to stall during long communication delays.

We propose AGILE, a lightweight asynchronous GPU-centric I/O library that eliminates deadlock risks and integrates a flexible HBM-based software cache. AGILE overlaps computation and I/O, improving performance by up to 1.88$\times$ across workloads with diverse computation-to-communication ratios. Compared to BaM on DLRM, AGILE achieves up to 1.75$\times$ speedup through efficient design and overlapping; on graph applications, AGILE reduces software cache overhead by up to 3.12$\times$ and NVMe I/O overhead by up to 2.85$\times$; AGILE also lowers per-thread register usage by up to 1.32$\times$.


\end{abstract}

\ccsdesc[500]{Information systems~Storage architectures}
\ccsdesc[500]{Computing methodologies~Parallel computing methodologies}
\ccsdesc[500]{Hardware~External storage}










\maketitle

\vspace{20pt}
\section{Introduction}

Graphics Processing Units (GPUs) have become the de facto accelerator widely used for computationally intensive applications such as graphics rendering~\cite{ren2021chopin, kilgard2012gpu}, deep learning~\cite{kilgard2012gpu, mittal2019survey}, and high-performance computing~\cite{wang2016gunrock, zhang2022egraph, wang2021grus}. However, modern applications are increasingly data-intensive, often processing data that far exceeds GPU memory capacity~\cite{gholami2024ai, maurya2024breaking, rajbhandari2021zero}. For example, training large-scale models like GPTs~\cite{achiam2023gpt,bi2024deepseek} involves billions of parameters and terabytes of training data~\cite{raiaan2024review}. Similarly, analyzing large graphs for social networking or ranking websites touches on billions of vertices and trillions of edges~\cite{ching2015one}. Recommender systems also handle data ranging from gigabytes to petabytes~\cite{raza2019progress}. Moreover, while GPUs' computational power has increased rapidly, their memory capacity has not kept the same pace~\cite{gholami2024ai}. These new trends necessitate innovative memory extension techniques and optimizations.

To expand GPUs' memory, existing solutions resort to CPU memory~\cite{Nvidia-unified-memory, zhang2023g10, allen2021depth, ren2021zero}. For example, Nvidia Unified Memory enables GPUs and CPUs to share a single memory address space so that GPUs can access CPU memory without explicit memory copies~\cite{Nvidia-unified-memory}. 
However, scaling the CPU memory to tens of terabytes is still considered a challenge~\cite{qureshi2023gpu}.
Another approach is extending GPU memory using SSDs~\cite{bae2021flashneuron, bergman2019spin, wu2024ssdtrain}, which provide much larger space but entail sophisticated designs for better performance. 
GPUDirect Storage~\cite{Nvidia_gpu_direct_storage} enables direct data transfers between GPUs and SSDs without involving the memory of the CPU as an intermediary, thereby eliminating the overhead of using CPU memory as a staging buffer. 
Microsoft proposes DeepNVMe~\cite{deepspeed_DeepNVMe}, which offers additional optimizations, such as asynchronous I/O operations and integration with ZeRO-Infinity~\cite{rajbhandari2021zero} for large neural networks. However, GPUDirect Storage and DeepNVMe still require the CPU to initiate the data transfer. As the computational workloads are offloaded onto the GPUs, the CPU lacks visibility to requests made by GPU threads in flight. Consequently, frequent synchronization between the GPU and the host CPU is necessary, leading to significant performance degradation~\cite{qureshi2023gpu}.

The emerging interconnect technology CXL is built on top of PCIe and offers new protocols, such as \texttt{CXL.memory} and \texttt{CXL.cache}, to efficiently extend host memory~\cite{cxl3-0}.
\texttt{CXL.memory} allows devices to use \texttt{load} or \texttt{store} instructions to access other devices' memory or storage.  \texttt{CXL.cache} further enables devices to coherently cache memory that physically resides on other devices. CXL-enabled SSDs are promising candidates for helping maintain coherence for memory expansion with SSDs~\cite{yang2023overcoming}, but are not currently a complete solution for expanding GPUs' memory. 
This is because the flash memory access time is at the microsecond level~\cite{yang2023overcoming}, which is orders of magnitude higher than High-Bandwidth Memory (HBM), where CXL is primarily deployed.  A solution to hide the latency of SSDs is still necessary. 

Overlapping memory access with computation is a common technique used to tolerate slow data movement~\cite{hijma2023optimization, chen2024centauri, wu2024ssdtrain}. For example, ALCOP~\cite{huang2023alcop} utilizes the CUDA-provided asynchronous data movement API to explore multi-stage pipelining.  This avoids GPU idle time caused by synchronous data movement.
However, inside a GPU kernel, only asynchronous data movement from global memory (or pinned host memory) to shared memory can be initiated using existing CUDA APIs~\cite{cuda-memcpy-async}, 
and the GPU's shared memory is limited, e.g., 164 KB per Streaming Multiprocessor on an A100 
GPU~\cite{gpu-shared-memory}. 
Using a larger buffer for asynchronous loads per thread has been demonstrated to have more performance benefits when using an overlapping technique~\cite{li2023performance}.


GPU-centric storage access is another method to avoid the synchronization overhead between GPUs and CPUs. BaM~\cite{qureshi2023gpu} is the first GPU-centric method that enables GPU threads to directly initiate NVMe I/O requests while bypassing the host CPU. It tolerates long SSD access latency via massive concurrent I/Os enabled by the GPU's high parallelism. This approach eliminates CPU intervention overhead. However, it adopts a synchronous access model, and threads must wait for the I/O requests to be completed before concurrently starting computation or issuing other commands. As a result, communication time cannot be hidden in each GPU thread, and applications must rely on runtime warp scheduling to preempt stalled warps and schedule other ready warps to avoid wasting GPU cycles~\cite{joseph2024wasp}, which is not always effective and leaves space for further optimization opportunities.

In contrast, an asynchronous I/O model can better tolerate long latency in accessing SSDs by overlapping communication with computation~\cite{lee2019asynchronous}. 
However, designing a GPU-centric asynchronous I/O model is challenging, as the massive GPU threads may compete on shared resources, e.g., NVMe queues, software-defined cache, etc., leading to performance degradation.
Using locks before accessing these shared resources is a common method to avoid resource conflicts, but in an asynchronous model, allowing threads to hold locks can lead to deadlock issues. For instance, if multiple threads asynchronously request SSD data, a request queue can fill prior to commands that check for completion and subsequently clear the completed request from the request queue, creating a deadlock. In addition, efficient lock handling is necessary to avoid performance degradation from the software API side.

Moreover, BaM~\cite{qureshi2023gpu} only supports a fixed cache policy for its software cache on GPUs' HBM.  
This limits the cache policy customization for various applications. As new caching policies~\cite{einziger2017tinylfu, pires2024learning, neglia2018cache} are continuously designed, it is important for storage systems to choose the best software-defined caching policy under various workloads and requirements~\cite{wang2010sopa}.

To address these needs and challenges, we propose AGILE, a GPU-centric GPU-SSD integration that enables GPU threads to issue NVMe requests asynchronously and efficiently while eliminating deadlock risks.

Our contributions are highlighted as follows:
%
\begin{itemize}
    \item We propose AGILE, enabling the GPU to issue NVMe commands asynchronously. To the best of our knowledge, AGILE is the first GPU-centric asynchronous I/O model.
    \item 
    We implement a robust lock-based asynchronous transaction mechanism,
 which allows GPU threads to issue NVMe commands asynchronously without holding any locks.  Our approach efficiently eliminates possible deadlocks and data hazards.
    \item We integrate a flexible software cache hierarchy in AGILE to utilize GPU HBM, which allows users to customize their cache policy and provides a simple interface for increased usability.
    \item We evaluate AGILE on micro-benchmarking and applications. The results show that AGILE enables overlapping at the thread level and achieves up to 1.88$\times$ speedup over a synchronous I/O model. Compared with state-of-the-art work, BaM, AGILE achieves up to 1.75$\times$ reduction in end-to-end execution time on DLRMs; in graph applications, AGILE demonstrates lower API overhead in managing software cache and NVMe I/O requests up to 3.12$\times$ and 2.85$\times$, respectively; furthermore, AGILE consumes fewer registers and exhibits up to 1.32$\times$ reduction in the usage of registers.
    \item We open-source AGILE with detailed guides for users to leverage AGILE and customize AGILE components in various applications: {\color{blue}\url{https://github.com/arc-research-lab/AGILE}}
\end{itemize}


\section{Background \& Design Challenges}
In this section, we first introduce the background of the NVMe protocol and how GPU threads are scheduled and hide memory access latency in CUDA. Then, we present the challenges in supporting an asynchronous I/O model on GPUs.
\subsection{Background of NVMe Protocol}
Non-Volatile Memory Express (NVMe) is a standard protocol that allows software to communicate with non-volatile memory via PCIe~\cite{nvme}. Software can access an NVMe SSD via an I/O queue pair, consisting of a submission queue (SQ) and a completion queue (CQ). 
With an I/O queue pair, the software is responsible for maintaining the SQ tail pointer, which indicates the next available SQ entry (SQE) for a new command, and the CQ head pointer, which is used to receive the next completion from the SSD.
To issue an NVMe command, the software writes a new command to the next available SQE and notifies the changes in SQ to the SSD by moving the SQ tail pointer and updating the new SQ tail by writing to the corresponding SQ doorbell register in the SSD's PCIe Base Address Registers (BAR). 
Then, the SSD fetches the newly added command, and after execution, the SSD returns a completion to the next available CQ entry (CQE). After receiving a completion, the software needs to respond to SSD by increasing the CQ head pointer and updating the associated CQ doorbell register. This is necessary for SSDs to release the CQE and reuse it for another command; otherwise, the SSDs will stall while waiting for available CQEs. This queue-based approach also allows software to issue multiple commands in a batch and increase the SQ tail pointer by the number of newly inserted commands.
The software can detect and process the completion message by either polling the CQ or responding to an interrupt triggered by the SSD. To achieve high parallelism, NVMe SSDs allow multiple SQs/CQs to be registered and used concurrently. 
\subsection{GPU Threads Scheduling \& Asynchronous Data Movement in CUDA}
To meet the increasing high throughput demands, modern GPUs can execute tens of thousands of threads in parallel via Single Instruction, Multiple Threads (SIMT)~\cite{ptx-instruction}. 
The GPU threads are grouped into thread blocks, and the threads in each thread block will be scheduled onto the same Streaming Multiprocessor (SM) \cite{NvidiaProgrammingGuide}. If the hardware resource, such as the number of registers and the shared memory, is enough for an SM to serve more than one thread block, each SM can accommodate multiple thread blocks simultaneously. 
Current GPUs adopt a static resource allocation model, which can cause SM underutilization. Once the thread blocks are scheduled onto SMs, they will occupy the SMs until their tasks are finished.
This prevents new thread blocks from being scheduled, even if the scheduled thread blocks are stalled due to some high-latency operations. 
This problem of SM underutilization is mitigated by warp-level scheduling. The SM will schedule threads at the granularity of warps (typically 32 threads in a warp). If some warps stall due to high-latency operations such as fetching data from memory or SSDs, other ready warps from the same thread block or different thread blocks can be scheduled to keep the SM busy. However, this mechanism is not sufficient, especially when many warps encounter memory or I/O stalls. 

To avoid stalls caused by memory access, users can use asynchronous data movement APIs such as \texttt{cuda::memcpy\_async}~\cite{cuda_async} or 
\texttt{cp.async}~\cite{ptx-instruction} in CUDA to hide latency with computation tasks. However, these asynchronous data movement APIs only allow data transfers from GPU global memory or pinned host memory to shared memory in SMs~\cite{cuda-memcpy-async}. 
Using larger buffers for asynchronous loads will lead to a higher performance increase~\cite{li2023performance}, but the shared memory is limited in each SM, e.g., 164 KB per SM on an A100 GPU~\cite{gpu-shared-memory}. 

\vspace{-5pt}
\subsection{Design Challenges in Asynchronous GPU-SSD integration}
\label{challenge}
\subsubsection{Deadlock in NVMe Queues}
\label{sec: deadlock}

Designing an efficient asynchronous model for GPU-SSD integration is challenging as a massive number of threads share limited resources such as NVMe queues and the software cache.
 Acquiring locks before accessing these resources is necessary to avoid conflicts, but can introduce deadlock.

For NVMe queues, when a thread puts a new NVMe command into an SQ, the corresponding SQ entry will remain locked to prevent other threads from using the same entry until the SSD has received the command. 
\begin{figure}[tbh]
\vspace{-5pt}
    \centering
    \includegraphics[width=0.9\linewidth]{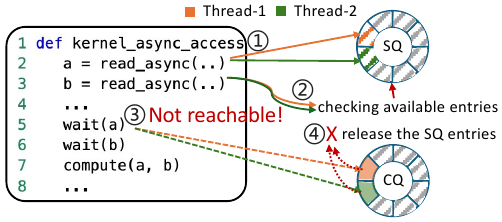}
    \vspace{-13pt}
    \caption{A deadlock example caused by sharing NVMe queues in asynchronous execution.}
    \label{fig:deadlock}
    \vspace{-5pt}
\end{figure}

Figure~\ref{fig:deadlock} illustrates an example of deadlock when Thread-1 and Thread-2 need to execute NVMe commands asynchronously in parallel. First, Thread-1 successfully acquires the SQ and places its read request into an available entry. However, before this thread can move to line 3, Thread-2 gains access to the SQ and adds its request to the last available entry, which fills the SQ~\textcircled{1}. Now, because the SQ is full, both threads become stuck at Line 3, they continue to check for the next available SQ entry~\textcircled{2}. 
Therefore, both threads cannot reach~\textcircled{3}, where they check the completions in CQ to confirm their issued commands have been processed by the SSD and then release locks in SQ.  
Even though the corresponding completions become available in the CQ, if Threads-1 and 2 own all the occupied SQ entries, none can be released~\textcircled{4}, resulting in a deadlock.  

For larger numbers of threads, this deadlock remains a concern as many threads will request multiple operands in 
line 3, hence, filling the queue prior to anyone reaching \textcircled{3}.

\vspace{-5pt}
\subsubsection{Deadlock in the Software Cache}
\label{sec:deadlock-SW-cache}
AGILE promises to offer flexibility in the software cache policy, and therefore, eliminating the potential for deadlock caused by the software cache is necessary. A common scenario resulting in a deadlock is simultaneous threads accessing multiple cache lines.  
For example, one compute kernel needs multiple operands that are stored in different cache lines. To prevent redundant SSD accesses, once a thread checks the software cache and the requested data is found---i.e., a cache hit occurs---%
access to the corresponding cache lines must be atomic to avoid eviction before accesses in process are completed.
When multiple threads
block cache line eviction while requesting new cache lines, a deadlock could occur.



\vspace{-5pt}
\subsubsection{Potential Performance Degradation}
\label{sec:performance}
Flash memory cannot be accessed randomly, and data is managed at a coarse-grained page level, typically 4KB per page~\cite{gal2005algorithms}. 
Therefore, the software cache line should align with the SSDs' granularity. This alignment can avoid redundant I/Os when multiple threads access different parts of the same SSD page concurrently. To ensure correctness during accessing the same cache line simultaneously, atomic operations are required to avoid conflicts and data hazards.
It is crucial to implement an efficient lock mechanism to prevent performance degradation and deadlock. 

Furthermore, in NVMe queues, although multiple threads can insert their commands into the same SQ concurrently, updating the SQ doorbell register must be serialized. This is because concurrent writes to the same doorbell registers may cause inconsistent SQ tail values in SSDs. Besides, the serialization ensures memory consistency so that the newly submitted commands are visible in global memory before the SQ doorbell registers are updated. Improper handling of this serialization may also cause performance degradation.

Lastly, real-world SSD devices only support a small number of I/O queue pairs compared to the massive living GPU threads. For example, a maximum of 128 queue pairs in Samsung 980 PRO NVMe SSD~\cite{samsung980pro-depth}. Therefore, the completions from SSDs tend to concentrate in a small number of completion queues, which requires an efficient and low-overhead mechanism to consume the completions to avoid stalls from SSDs.

\vspace{-5pt}
\section{AGILE Design \& Implementation}

In this section, we will first give an overview of AGILE in Section~\ref{overview}. Then, we present the main components of AGILE. In Section~\ref{agile-service}, we will discuss how AGILE avoids the deadlock problem resulting from NVMe queues and processes completions from SSDs in parallel. We will discuss how AGILE deals with the serialization process required by the NVMe SQs and coalescing redundant requests in Section~\ref{sec:issue}. In Section~\ref{cache}, we will present the software-managed cache in AGILE and discuss how AGILE extends cache coherency to user-specified buffers. Finally, we will present an example program, illustrating how AGILE can be used, and introduce a debug option provided in AGILE. 

\vspace{-5pt}
\subsection{Overview of AGILE System}
\label{overview}

\begin{figure}[tbh]
    \vspace{-10pt}
    \centering
    \includegraphics[width=0.95\linewidth]{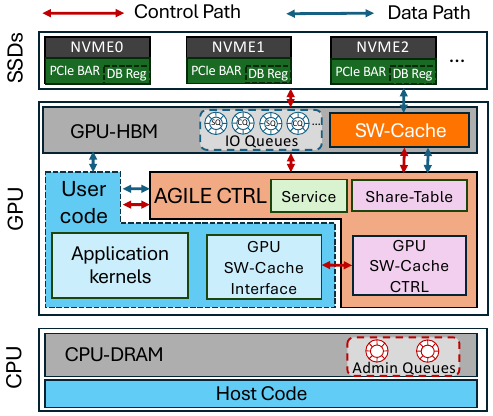}
    \vspace{-10pt}
    \caption{Overview of system architecture adopting AGILE.}
    \label{fig:agile-framework}
    \vspace{-8pt}
\end{figure}

Figure~\ref{fig:agile-framework} presents an overview of the AGILE system, which enables efficient asynchronous GPU-SSD communication. 
The system involves three types of hardware, including SSDs, a GPU, and a host CPU. 
The host CPU manages admin queues, located in DRAM, to establish GPU-SSD PCIe peer-to-peer (P2P) communication. The NVMe SSDs are connected to the system via PCIe, their PCIe BARs are exposed to the host CPU for management, and their doorbell registers are registered to GPU for GPU-centric data transfers. 
Within the GPU, AGILE consists of a lightweight service to 
handle I/O queues for users (Section~\ref{agile-service}), a software controller to manage cached data in HBM (Section~\ref{cache}), and a Share Table to extend cache coherency to user-specified buffers (Section~\ref{share-table}). Users can interact with AGILE through the AGILE controller (AGILE CTRL), which provides simple APIs for requesting or accessing data in SSDs or the software cache.

To establish the PCIe P2P communication, the SSDs and the GPU must be able to access the other device's memory. To let an NVMe SSD access I/O queues (SQs/CQs) and the software cache, we need to allocate a contiguous memory space on GPU HBM, pin the memory space to avoid being swapped out, and get the physical address of the memory space to enable Direct Memory Access (DMA) for the SSD to access the GPU HBM.
GDRCopy~\cite{gdrcopy} is designed for direct GPU memory access from third-party devices.  
It runs in kernel space and serves userspace calls for allocating and pinning contiguous memory on the GPU. 
We modify the GDRCopy kernel module and invoke \texttt{nvidia\_p2p\_put\_pages} in the kernel space, enabling userspace applications to access the mapping table that translates virtual addresses into physical addresses of GPU memory.
Then, the physical addresses of SQs/CQs are registered to SSDs via the admin queues on the host CPU. 
To let the GPU notify NVMe SSDs after generating new commands, we use memory-mapping
(\texttt{mmap}) to expose the SSDs' PCIe BAR to userspace and then register the doorbell registers to the GPU using \texttt{cudaHostRegister} with the \texttt{cudaHostRegisterIoMemory} flag. After this initialization process, the GPU threads can insert NVMe commands to SQs in HBM and update the doorbell registers to notify the NVMe SSDs directly, and the NVMe SSDs are able to fetch commands in GPU HBM, process them, and update the completion messages to CQs in GPU HBM directly.
In AGILE, the initialization process requires CPU intervention and must be performed at the beginning of the program using AGILE.

 
 


AGILE provides two types of asynchronous APIs and an array-like synchronous API. 
The asynchronous API \texttt{prefetch(src)} is used to issue data requests from SSDs to the GPU software cache, and then the user threads can access the data directly in the GPU software cache.
Another asynchronous API, \texttt{async\_issue(src,dst)}, is similar to \texttt{cuda::memcpy\_async}~\cite{cuda_async} or 
\texttt{cp.async}~\cite{ptx-instruction} in CUDA, but the \texttt{src} and \texttt{dst} in AGILE are more flexible and can be either SSDs' addresses or user-specified buffers in GPUs' global memory.
By using user-specified buffers with \texttt{async\_issue(src,dst)}, GPU threads can save multiple data chunks for later use safely without holding locks in the software cache, thereby avoiding the deadlocks described in Section~\ref{sec:deadlock-SW-cache}.
However, the increased flexibility of \texttt{src} and \texttt{dst} in \texttt{async\_issue(src,dst)} may introduce data hazards, and we will present our solution in Section~\ref{share-table}. 
The \texttt{async\_issue(src,dst)} will return a barrier to let the user threads know if the data transfer is completed. Lastly, the array-like synchronous API allows users to simply view the SSDs as a two-dimensional array, and AGILE automatically checks the software cache and issues requests if the data is not available.
\subsection{AGILE Service}
\label{agile-service}

As mentioned in Section~\ref{challenge}, allowing threads to hold locks on NVMe queues is risky and can cause deadlock. However, locking SQs is necessary so that commands do not collide.
To address this problem, we propose a lightweight AGILE service that runs in the background on the GPU and interacts with user threads.

\subsubsection{Avoid deadlock from NVMe queues}
\label{avoid-deadlock}
To eliminate the deadlock risk, AGILE creates a lightweight kernel daemon on the GPU to keep checking completion queue entries (CQE) for all registered NVMe CQs in a non-blocking fashion. 
This service frees the user threads from the burden of processing completion messages and automatically releases shared resources for user threads after completion. Once the AGILE service receives a completion from the CQs, the corresponding locks in SQs will be released.  This allows additional SQ requests to proceed and avoids deadlock even when user threads issue multiple request commands.




\begin{figure}[tbh]
\vspace{-10pt}
    \centering
    \includegraphics[width=1\linewidth]{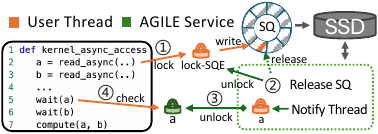}
    \vspace{-20pt}
    \caption{Avoiding NVMe Queue Deadlocks in AGILE. 
    }
    \label{fig:exchange-lock}
    \vspace{-5pt}
\end{figure}

Figure~\ref{fig:exchange-lock} illustrates the process of how the AGILE service assists user threads in issuing commands. In Figure~\ref{fig:exchange-lock} line 2, when a user thread successfully locks the SQ, it can safely enqueue the command into the SQ entry~\textcircled{1}. Then, it will handoff \texttt{lock-SQE} to the AGILE service and receive back a barrier (lock \texttt{a}) representing the status of the transaction.  
Thus, when a thread reaches lines 2--3 and cannot add its requests to the SQ because it is full, once the AGILE service receives completions from SSDs, it can release the appropriate SQ entry directly and then clear the appropriate transaction lock~\textcircled{2} so that the user thread will not be blocked forever. Meanwhile, the AGILE service will notify the corresponding barrier by clearing the lock \texttt{a}~\textcircled{3} to indicate that the transaction is finished.
Finally, in line 5,
if the thread arrives at line 5 prior to the SSD access completion, it will wait for the AGILE service to clear the lock \texttt{a}~\textcircled{4}.

Since the completions may be returned out of order relative to the issued commands, the AGILE service tracks the mapping between each completion and its corresponding SQE via the Command Identifier (CID), which is a 16-bit field in the NVMe command and should be unique to identify commands within a batch using the same SQ. 
\subsubsection{Polling Completion Queues (CQs)} 
Processing CQs efficiently is critical to sustain high throughput in a GPU-centric asynchronous I/O model.
In practice, the NVMe SSDs only support a limited number of CQs. For example, the Samsung 980 PRO NVMe SSD supports up to 128 CQs~\cite{samsung980pro-depth}. In contrast, GPU applications typically involve a great number of threads, many of which need to share the same CQ. As a result, completions may tend to concentrate in a small number of CQs, which could lead to contention and performance bottlenecks. To ensure timely completion processing, AGILE increases intra-CQ polling parallelism by adopting a warp-centric CQ polling strategy, where each warp concurrently processes 32 CQEs within a CQ at every iteration. Meanwhile, AGILE only uses a small number of warps for CQ polling and rotates across all registered CQs in a round-robin fashion.

\begin{algorithm}
\caption{Warp-centric CQ polling}
\label{alg:polling}
\begin{algorithmic}[1]  
\Function{CQ\_Polling}{$cq\_idx$}
    \State $offset, mask, phase\_bit \gets load\_CQ(cq\_idx)$
    
    \If{$mask[warp\_idx] == 0$}
        \State $pos \gets offset + warp\_idx$
        \State $valid \gets process\_CQE(cq\_idx, pos, phase\_bit)$
        \State $mask[warp\_idx] \gets valid$
    \EndIf
    \If{$mask == 0xFFFFFFFF$}
        \State $mask \gets 0$
        \State $update\_CQ(cq\_idx, offset)$
    \EndIf
    \State $update\_mask(cq\_idx, mask)$
    
\EndFunction
\end{algorithmic}
\end{algorithm}

Algorithm~\ref{alg:polling} describes the warp-centric CQ polling routine used in the AGILE service to process CQs efficiently. When invoked, the warp is assigned with a specific CQ, and each thread is responsible for checking a single CQE within a 32-entry window.
In the warp-centric CQ polling service, the threads first load the current polling offset, the CQ phase bit for monitoring the changes in CQEs, and a 32-bit mask that represents the completion status of the CQEs (line 2). If the corresponding bit in the mask is unset, which indicates the completion is not received, the threads will compare the CQEs' phase bit with the expected value. If new completions are found, the associate bits in the mask will be set to 1 (lines 5-6). When all threads in the warp detect valid completions, indicated by the mask being fully set, the polling service considers the window fully processed. If the window is fully processed, the warp will update the CQ doorbell register to notify the SSD and reset the mask for the next round (lines 9-10). The mask will be updated each time to save the current status of the target CQ (line 12). This warp-coordinated approach increases the parallelism in processing each CQ while minimizing the divergence across threads in a warp because all threads operate on physically contiguous CQEs and follow the same polling logic.

\subsection{AGILE Request Issuing Mechanism}
\label{sec:issue}
 As discussed in Section~\ref{sec:performance}, the SQs require an efficient serialization mechanism before updating the SQ doorbell registers to avoid performance degradation.
In this subsection, we first present how user threads issue NVMe commands. Then, we illustrate how AGILE coalesces redundant requests at the warp level.
\vspace{-5pt}
\subsubsection{Serialization process in NVMe SQs}
\label{sec:serialization}
In AGILE, each SQE is associated with a lock that can have three possible states: \texttt{EMPTY}, \texttt{UPDATED}, and \texttt{ISSUED}.
Algorithm~\ref{alg:serialization} illustrates the serialization process for issuing NVMe commands. When a user thread needs to issue an NVMe command, it first selects an SQ associated with the target SSD based on its thread index and attempts to submit the command to this SQ if it has an available SQE for a new command (line 2). If the SQ is full, the thread will try to submit commands to another SQ by simply increasing the index of the target SQ. After enqueuing the commands to an SQ (line 6), AGILE sets the state of the corresponding SQE's lock to \texttt{UPDATED}, which indicates the command is now visible in the global memory and can be safely notified to the SSD. To ensure the SSD is properly notified, all threads will attempt to update the associated SQ doorbell register and verify whether their commands have been issued (line 9). A thread that successfully acquires the lock for the SQ doorbell register increases the SQ tail (line 15), during which it scans the SQEs in order and updates the SQEs' states from \texttt{UPDATED} to \texttt{ISSUED}. This process continues until it encounters an SQE in the \texttt{EMPTY} state, which either marks the end of the current batch of commands or indicates that the corresponding SQE is not visible in the global memory yet. Then, this thread will update the SQ doorbell register and release the lock (line 15). Finally, all threads verify the states of their respective SQEs (line 17) to confirm if the commands have been successfully issued to the SSD. 
Once the completions are received by the AGILE service, the corresponding SQEs' states are reset to \texttt{EMPTY}, allowing them to be reused for future commands.
\begin{algorithm}
\caption{Serialization process in SQs}
\label{alg:serialization}
\begin{algorithmic}[1]  
\Function{Attempt\_Enqueue}{$sq\_idx, cmd$}
    \State $sqe = check\_full(sq\_idx)$
    \If{$sqe == -1$}
    \State \Return \textbf{false}
    \EndIf
    \State $enqueue\_cmd(sq\_idx, sqe, cmd)$
    \State $update\_SQE(sq\_idx, sqe, ENQUEUE)$
    \Repeat
        \State $status \gets Attempt\_SQDB(sq\_idx, sqe)$
    \Until{$status == SUCCESS$}
    \State \Return \textbf{true}
\EndFunction

\Function{Attempt\_SQDB}{$sq\_idx, sqe$}
    \If{$acquire\_lock(sq\_idx)$}
        \State $move\_SQ\_tail(sq\_idx, sqe)$
    \EndIf
    \State \Return $check\_SQE(sq\_idx, sqe)$
\EndFunction

\end{algorithmic}
\end{algorithm}
\vspace{-10pt}
\subsubsection{Coalescing identical requests at the warp level} 
To avoid redundant requests, AGILE coalesces identical data requests issued by different threads, which is essential because user threads may independently request the same data chunk from SSDs. 

For \texttt{prefetch()} and the array-like interface, 
AGILE employs a two-level coalescing strategy.
The first level occurs at the warp level, where CUDA warp-level primitives~\cite{cuda-warp} are used to examine duplicate requests.
Then, AGILE selects one thread to forward the request to the second-level coalescing stage. 
The second level is handled by the AGILE software cache (Section~\ref{cache}), which filters remaining redundant requests that are not eliminated in the first warp-level coalescing stage. AGILE prioritizes the warp-level coalescing since accessing the shared software cache requires atomic operations to maintain consistency, which creates critical sections and serializes execution. This serialization can cause stalls and different execution paths for threads in a warp, which introduces warp divergence and degrades overall GPU performance.

For \texttt{async\_issue(src,dst)}, which mimics \texttt{cp.async}~\cite{ptx-instruction} or \texttt{cuda::memcpy\_async}~\cite{cuda_async} in CUDA and no warp-level coalescing is performed.
Even if threads in a warp request the same data, each thread will still obtain its own copy of the requested data. 
Therefore, in AGILE, the redundant requests are only coalesced at the software cache level, and AGILE delegates the warp level optimization to users. Moreover, \texttt{async\_issue(src,dst)} provides more flexibility compared to the CUDA APIs, which can introduce potential data hazards. These data hazards are addressed through the Share Table mechanism, which will be described in Section~\ref{share-table}.

\subsection{AGILE Software Cache}
\label{cache}
A software-managed cache can significantly reduce SSD I/O traffic by storing frequently accessed SSD data on the device~\cite{qureshi2023gpu, liu2017hardware, hong2024bandwidth}. AGILE also enables this feature and provides built-in cache policies as well as interfaces for users to customize cache policies.
In AGILE, all SSD data accesses are routed through the software cache to ensure coherency and to coalesce the redundant SSD requests.

In AGILE, each cache line has four possible states: \texttt{INVALID}, \texttt{BUSY}, \texttt{READY}, and \texttt{MODIFIED}. 
When user threads request any data, AGILE first checks the user-specified cache policy and obtains the target cache line index. 
There will be 4 possible cases:
\textbf{(a) cache hit and data is valid.} 
If the state of the cache line is 
\texttt{READY}, or \texttt{MODIFIED}, it means the data is already in GPU HBM, and the threads can directly obtain the requested data. \textbf{(b) cache miss and no eviction required.} In this case, the state is \texttt{INVALID}, and the thread will issue an NVMe command to load data from SSD to HBM and change the cache line state to \texttt{BUSY}.
\textbf{(c) cache hit, but the data is invalid.} This happens when the cache line state is \texttt{BUSY}. This means the data has already been requested by another thread, and this thread will either wait (synchronous APIs) or append its buffer to the corresponding linked list. \textbf{(d) cache miss and eviction required.} This occurs when the cache line is reserved, and the state is not \texttt{INVALID}. Then, AGILE will trigger a cache line eviction if the cache line state is \texttt{READY}, \texttt{MODIFIED}, or \texttt{BUSY}. AGILE will simply reset the cache line if the state is \texttt{READY}, and write \texttt{MODIFIED} cache line to the SSDs and change the state to \texttt{BUSY}.
If the state is \texttt{BUSY}, the corresponding cache line cannot be evicted until the processing is finished, and AGILE will let the user-specified GPU software cache policy decide whether to wait or find another cache line.


\subsubsection{Extending Coherency to User-specified Buffers}
\label{share-table}
It is worth noting that \texttt{async\_issue(src,dst)} in AGILE is conceptually similar to \texttt{cuda::memcpy\_async}~\cite{cuda_async} or 
\texttt{cp.async}~\cite{ptx-instruction} in CUDA, which enables asynchronous data movement to hide memory latency. 
However, the CUDA's asynchronous APIs are limited to specific memory paths, i.e., transferring data from the global memory or pinned host memory into the shared memory on the SM~\cite{cuda-memcpy-async}. In contrast, \texttt{async\_issue(src,dst)} in AGILE provides greater flexibility in the source and destination addresses, both of which can be SSD data or user-specified GPU buffers. This enhanced flexibility, however, introduces potential data hazards. 
For example, a thread may issue an \texttt{async\_issue(src,dst)} to fetch data from SSD directly to the user-specified buffer, while other threads can concurrently access the same data from the AGILE software cache. 
If the user-specified buffer or software cache is modified without coordination, data hazards such as read-after-write (RAW), write-after-read (WAR), and write-after-write (WAW) can occur, where threads may observe stale or partially updated data.

To address these data hazards, AGILE provides a compile-time option to enable the user-specified buffers to be integrated into the AGILE software cache and safely shared among multiple user threads.
If enabled, by default, AGILE will maintain a hashtable-based Share Table to track user-specified buffers' ownership and apply a software-managed coherency protocol inspired by the MOESI model~\cite{sweazey1986class} to ensure consistency across different access paths.

Unlike the original MOESI model, where each thread maintains its own copy of data, AGILE maintains the coherency by sharing the pointers to the user-specified buffers, which allows all threads to access the same physical memory region. This eliminates redundant data duplication and avoids extra copies between threads. In AGILE, the MOESI is reinterpreted to reflect the relationship and responsibility between user threads and their shared buffers. 
Specifically, when a thread requests data for its buffer, the thread receives exclusive ownership of that buffer. Meanwhile, the Share Table records the source of the data in the buffer and stores the pointer to this buffer. When other threads request the same source of data, the Share Table will return the existing pointer to that buffer and increment a corresponding reference counter of the shared buffer to indicate the usage. If any threads attempt to modify the buffer, the buffer will switch to the Modified State, and the original owner of the buffer will be responsible for propagating the updates to L2 cache -- software cache in GPU HBM -- after other threads finish using the buffer.

When this Share Table is enabled, it will have the highest priority in the AGILE software cache hierarchy. When new requests arrive, AGILE will first consult the Share Table to determine if any user thread owns a valid buffer of the requested data. If no record is found, AGILE will fall back to the software cache or issue a new request to the SSD and register this buffer in the Share Table. Similar to the flexible customization in software cache, AGILE allows users to design their own sharing policy and integrate it into AGILE seamlessly to meet various application needs.

\vspace{-5pt}
\subsection{AGILE Abstraction and Software APIs}

Listing~\ref{lst:agile-example} shows an example GPU program that uses AGILE. Users can define their software cache policy (line 1) or directly choose the built-in software cache policies and specify the software cache policy in line 2. To provide flexibility in software cache and share table policies, AGILE employs the curiously recurring template pattern (CRTP) to implement the software cache and share table control logic. CRTP enables compile-time polymorphism and avoids using virtual functions. The software cache and the Share Table policies are specified in line 2. 

Because AGILE allows users to provide customized policies, where processing on locks is necessary and may introduce new deadlock risks, AGILE provides a debug option at compile time to track acquired locks within each thread using a lock chain implemented as a linked list (line 6). If this debug option is enabled, when a thread tries to acquire a target lock but fails, it will scan all previously acquired locks and mark these acquired locks are dependent on the target lock to release. Then, it will check if any acquired lock exists in the dependency chain of the target lock -- if a circular dependence results in a deadlock. If a circular dependency happens, AGILE will report it to users. 

Lines 8 - 19 present the three methods to access SSDs in AGILE. Line 9 is an example of the \texttt{prefetch()}, which asynchronously loads the data from a target SSD to the software cache. Line 12 shows how users can register a user-specified buffer to AGILE and use \texttt{async\_issue(src,dst)} to load or store data asynchronously (lines 13 - 15). 
For \texttt{asyncRead()}, users need to verify if the transfer is completed before using (line 14), while the  \texttt{asyncWrite()} will ensure the data is updated to the software cache and the write command is issued, and the buffer is available right away for other purposes.
AGILE also provides an array-like synchronous API that views the SSDs as a two-dimensional array, where the first dimension specifies the SSD indices and the second dimension is the data position to access (lines 18 - 19).
\begin{table}[tbh]
\begin{minipage}{\columnwidth}
\begin{lstlisting}[caption={Example GPU program using AGILE.}, label={lst:agile-example}]
class GPUCache:public GPUCacheBase<GPUCache>{...};
#define AGILE_CTRL AgileCtrl<GPUCache, ShareTable>
__global__ 
void kernel(AGILE_CTRL * ctrl, void * data){
  ...
  AgileLockChain chain;

  // Method-1: AGILE prefetch
  ctrl->prefetch(dev_idx, blk_idx, chain);

  // Method-2: AGILE async_issue
  AgileBufPtr buf(data + tid * ctrl->line_size);
  ctrl->asyncRead(dev_idx, blk_idx, buf, chain);
  buf.wait();
  ctrl->asyncWrite(dev_idx, blk_idx, buf, chain);

  // Method-3: AGILE array-like synchronous API
  auto agileArr = ctrl->getArrayWrap<int>(chain);
  int val = agileArr[dev_idx][idx];
}

int main(int argc, char** argv){
    // GPU Configurations
    AGILE_HOST host(...); 
    // Policy Configurations
    SHARE_TABLE_IMPL s_table(...); 
    GPU_CACHE_IMPL g_cache(...);
    host.setGPUCache(g_cache);
    host.setShareTable(s_table);
    // Add and open target SSDs in the program
    host.addNvmeDev("/dev/AGILE-xxx", ...);
    host.addNvmeDev("/dev/AGILE-xxx", ...);
    host.initNvme();
    // Initialize AGILE controller
    host.initializeAgile(...);
    // CUDA kernel parallelism configurations
    host.configKernelParallelism(...);
    host.queryOccupancy(kernel);
    // Start the lightweight AGILE service
    host.startAgile();
    // Execute the CUDA kernel
    host.runKernel(kernel, args...); 
    // Stop AGILE service
    host.stopAgile();
    // Close the opened SSDs
    host.closeNvme();
}
\end{lstlisting}
\end{minipage}
\end{table}

Lines 22 - 47 demonstrate the AGILE host-side code executed on the CPU. At Line 24, users specify the GPU configurations, e.g., selecting which GPU to use for the program. Lines 26 - 29 handle the initialization of AGILE's GPU software cache and share table policies. AGILE allows multiple NVMe SSDs to be configured and used simultaneously in the program, as shown in Lines 31 - 33. To utilize SDDs with AGILE, the devices must be bound to the AGILE-provided NVMe SSD driver, which creates a device file, \texttt{/dev/AGILE-NVMe-\$\{PCIe-BDF\}}, for each SSD. 
AGILE supports customized NVMe queue configurations for users to enable prioritization control across SSDs. 
At Line 35, AGILE allocates physically contiguous memory on HBM for NVMe I/O queues and registers these queues to the SSDs. 
Lines 37 - 38 configure the application kernel's launch configurations (i.e, \texttt{gridDim}, \texttt{blockDim}), compile the application kernel, and report the maximum number of active blocks per SM. The AGILE lightweight runtime service, described in Section~\ref{agile-service}, must be started (Line 40) and properly terminated (Line 44) before and after kernel execution (Line 42). Finally, the opened SSDs need to be closed at Line 46.

}
\section{Evaluation}
In the experiments, we first use a micro-benchmark to demonstrate the advantages of the asynchronous model over a synchronous model under different workload characteristics.
Then, we evaluate the scalability of AGILE using 4KB random read and write on various numbers of SSDs. 
To demonstrate the usability of AGILE, we compare AGILE with the state-of-the-art work BaM on Deep Learning Recommendation Models (DLRMs) and use various configurations. We further evaluate the API overhead of AGILE against BaM on graph applications to demonstrate AGILE's efficiency. Lastly, we report the pre-thread register usage of AGILE and BaM, which shows that AGILE is more lightweight in terms of GPU resource consumption.
\vspace{-5pt}
\subsection{Experimental Setup}
We evaluate AGILE on a Dell R750 server running Ubuntu 20.04, equipped with an Nvidia RTX 5000 Ada GPU~\cite{rtx5000ada}, a Dell Ent NVMe AGN MU AIC 1.6TB SSD~\cite{ssd_spec}, and two Samsung 990 PRO 1TB SSDs~\cite{990pro}.
The GPU and SSDs are attached to the server via PCIe Gen4x16 and Gen4x4, respectively. 
The Nvidia Driver 550.54 and the CUDA 12.8 are installed on the server for experiments. The modified Linux kernel drivers used in AGILE are tested on Linux 5.4.0-200-generic. 
\subsection{Comparison between asynchronous I/O and synchronous I/O}

We first demonstrate how AGILE's asynchronous I/O model enables overlapping between computation and communication to reduce end-to-end execution time. In this experiment, 1024 threads within a block are launched to issue 64 NVMe commands and perform computation on the returned data. In the synchronous mode, computation begins only after all data has been fetched. In contrast, the AGILE asynchronous mode enables computation and communication overlapping at the thread level.
Ideally, when computation and communication perfectly overlap with each other, the speedup can be defined by Equation~\ref{eq:idea_speedup}:
\vspace{-5pt}
\begin{equation}
\label{eq:idea_speedup}
\text{Ideal Speedup} = 
\begin{cases}
1 + \text{CTC}, & 0 \leq \text{CTC} \leq 1 \\
1 + \dfrac{1}{\text{CTC}}, & \text{CTC} > 1
\end{cases}
\end{equation}

\begin{figure}[tbh]
\vspace{-5pt}
    \centering
    \includegraphics[width=1\linewidth]{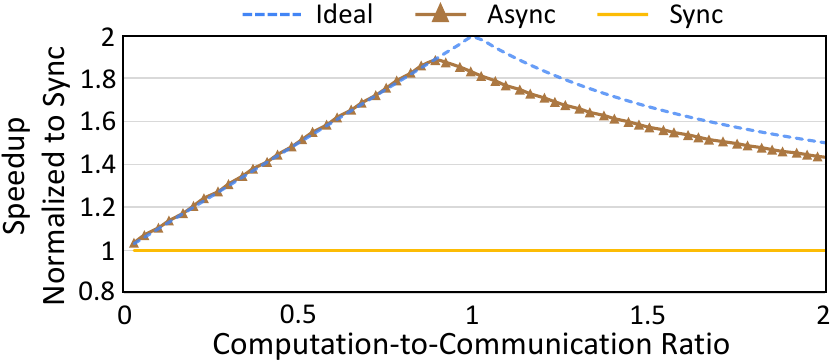}
    \vspace{-20pt}
    \caption{Speedup comparison of asynchronous I/O over synchronous I/O on workloads with different Computation-to-Communication Ratio (CTC).}
    \label{fig:ctc}
\end{figure}

As shown in Figure~\ref{fig:ctc}, we illustrate the effectiveness of AGILE’s thread-level asynchronous model by varying the computation-to-communication (CTC) ratio from 0 to 2 by increasing the number of computation iterations. AGILE asynchronous version can achieve up to 1.88x improvement over the synchronous baseline. The observed speedup increases with CTC until it reaches a peak where CTC is close to 0.9 and then gradually decreases when CTC further increases, which aligns with the theoretical trend.
The peak speedup occurs below CTC equals 1 because certain portions of the asynchronous pipeline stages, such as prefetching and the issuing logic, cannot be fully hidden by either computation or communication, which limits the ideal overlap. The experimental results demonstrate that AGILE's asynchronous I/O model is effective in hiding communication time, especially when the computation and communication are balanced.


\begin{figure}[tbh]
\vspace{-10pt}
    \centering
    \includegraphics[width=1\linewidth]{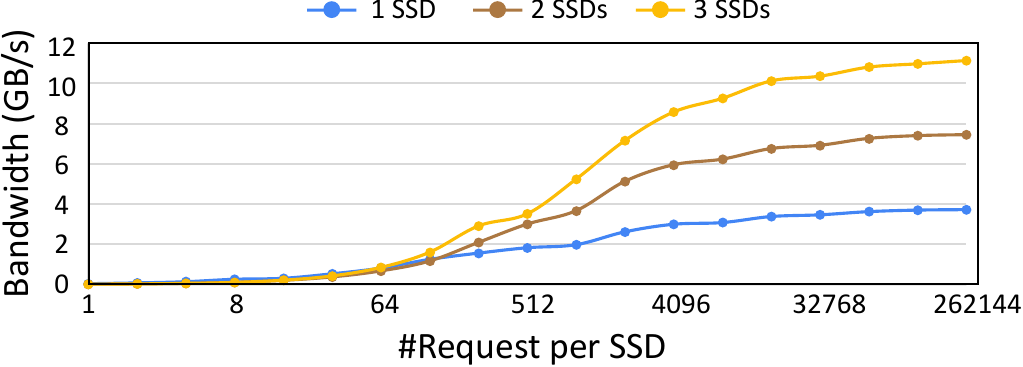}
    \vspace{-20pt}
    \caption{AGILE 4KB random read on multiple SSDs}
    \label{fig:rand-read}
    \vspace{-10pt}
\end{figure}

\begin{figure}[tbh]
    \centering
    \includegraphics[width=1\linewidth]{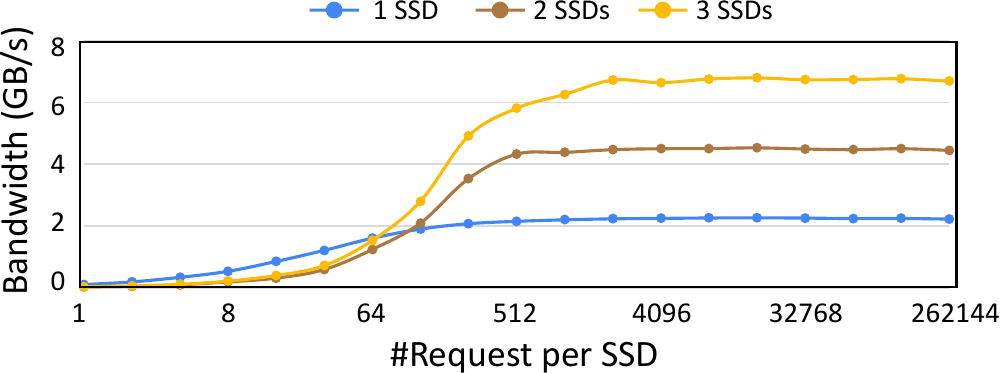}
    \vspace{-20pt}
    \caption{AGILE 4KB random write on multiple SSDs}
    \label{fig:rand-write}
\vspace{-10pt}
     
\end{figure}

\subsection{AGILE 4KB random read and write on multiple SSDs}

We evaluate the scalability of AGILE using 4 KB random read and write using 1, 2, and 3 SSDs, as shown in Figure~\ref{fig:rand-read} and Figure~\ref{fig:rand-write}, respectively. For experiments with more than one SSD, different SSDs are accessed in an interleaved manner. For example, requests 0, 2, 4, etc. are issued to SSD1, while requests 1, 3, 5, etc. are directed to SSD2.
In both 4 KB random read and write, AGILE exhibits scalable performance as the number of requests increases and can leverage multiple SSDs effectively.
For 4KB random reads in Figure~\ref{fig:rand-read}, the aggregate bandwidth saturates at 3.7 GB/s, 7.4 GB/s, and 11.1 GB/s with 1 SSD, 2 SSDs, and 3 SSDs, respectively, after approximately 32K concurrent requests per device.  
Figure~\ref{fig:rand-write} depicts the aggregate write bandwidth achieved by AGILE in the 4 KB random write workload, and AGILE saturates at 2.2 GB/s, 4.4 GB/s, and 6.7 GB/s with 1 SSD, 2 SSDs, and 3 SSDs, respectively.

\subsection{Evaluation on DLRM inference}

We further evaluate AGILE against BaM~\cite{qureshi2023gpu} on Deep Learning Recommendation Model (DLRM) inference. We use the Criteo 1TB Click Logs dataset~\cite{criteo-dataset} and construct the categorical feature vocabulary using the first three days of data.
To ensure consistent and efficient computation across all experiments, we use cuBLAS~\cite{cublas} for matrix multiplications. 
BaM and AGILE are used to fetch embedding vectors to HBM, and their kernels are integrated into the CUDA stream pipeline with cuBLAS kernels.
We keep the same clock replacement cache policy~\cite{corbato1968paging} and set the software cache size to 2GB for all experiments unless otherwise specified. For NVMe I/O queue configurations, we use 128 queue pairs, and the queue depth of each queue is set to 256 by default across all experiments unless otherwise specified.
We use AGILE in both the synchronous mode (AGILE sync) and the asynchronous mode (AGILE async). For AGILE sync and BaM implementation, we request data and perform computation on the requested data within the same epoch. For AGILE async, we prefetch data for the next epoch to enable overlapping of communication and computation. 

We adopt DLRM architecture from ~\cite{naumov2019deep} and evaluate several variants. In addition to projection layers (for dimensional alignment in matrix multiplication) and activation layers, the bottom MLP in Config-1 has three matrix multiplication kernels with dimensions 512-512-512, and the top MLP consists of three layers with sizes of 1024-1024-1024. Config-2 reduces the number of matrix multiplications to one in both the bottom MLP and the top MLP to represent a less computationally intensive model. In Config-3, we repeat the matrix multiplications six times to emulate a more computationally intensive workload. In all configurations, we measure the end-to-end execution time using a batch size of 2,048 and an epoch size of 10,000.

\begin{figure}[tbh]
\vspace{-8pt}
    \centering
    \includegraphics[width=1\linewidth]{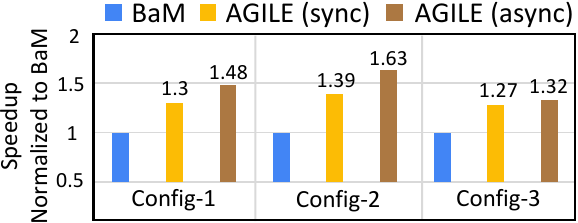}
    \vspace{-15pt}
    \caption{Speedup comparison of AGILE (async and sync modes) over BaM on different recommendation models.}
    \label{fig:dlrm-model}
    \vspace{-8pt}
\end{figure}

Figure~\ref{fig:dlrm-model} illustrates the speedup comparison of AGILE in both synchronous and asynchronous modes relative to BaM across three DLRM configurations. AGILE sync shows consistent improvement over BaM, achieving speedups of 1.3$\times$, 1.39$\times$, and 1.27$\times$ in Config-1, Config-2, and Config-3, respectively. The AGILE async further improves the performance by overlapping data movement with computation and reaches 1.48$\times$, 1.63$\times$, and 1.32$\times$ speedups in the same configurations.


\begin{figure}[tbh]
    \vspace{-10pt}
    \centering
    \includegraphics[width=1\linewidth]{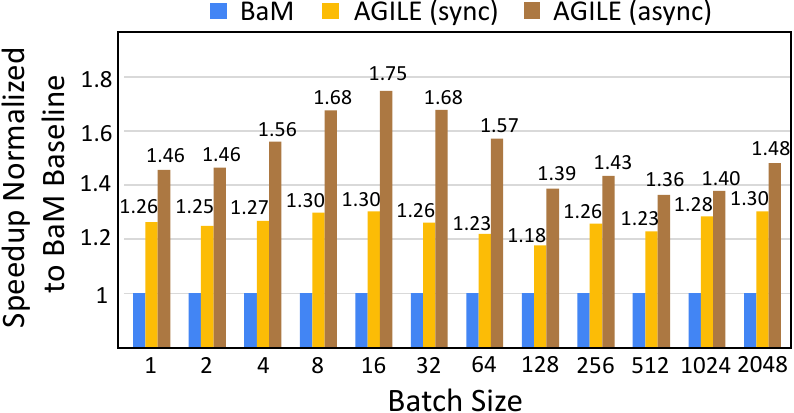}
    \vspace{-20pt}
    \caption{Speedup comparison of AGILE (async and sync modes) and BaM across varying batch sizes in DLRM inference.}
    \label{fig:dlrm-batch}
    \vspace{-15pt}
\end{figure}

To understand how AGILE performs under different workload granularities, we evaluate AGILE's speedup across a wide range of batch sizes using DLRM Config-1, which assesses the scalability of AGILE and BaM. Figure~\ref{fig:dlrm-batch} depicts the speedup of AGILE in sync and async modes normalized to the BaM baseline across batch sizes ranging from 1 to 2048. AGILE sync mode shows stable gains over BaM with speedup from 1.18$\times$ to 1.30$\times$. AGILE async also consistently outperforms AGILE sync across all batch sizes and reaches the peak speedup to 1.75$\times$ at a batch size of 16. These results demonstrate AGILE's ability to overlap computation and computation at scale. The results also indicate that the AGILE async benefits more when the batch size is smaller and near 16 in this DLRM inference, where the opportunity to hide communication is more significant.

\begin{figure}[tbh]
\vspace{-8pt}
    \centering
    \includegraphics[width=1\linewidth]{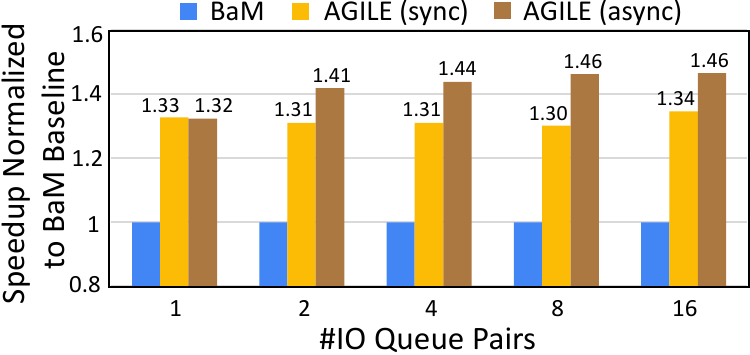}
    \vspace{-20pt}
    \caption{Speedup comparison of AGILE (async and sync modes) and BaM under varying numbers of I/O queue pairs in DLRM inference.}
    \label{fig:dlrm-queue}
    \vspace{-10pt}
\end{figure}

We further study the sensitivity of NVMe queue settings for both AGILE and BaM using DLRM Config-1 and a batch size of 2048. Specifically, we reduce the queue depth to 64 and sweep the number of queue pairs from 1 to 16, which introduces greater contention in the NVMe queues. Figure~\ref{fig:dlrm-queue} demonstrates that both AGILE sync and async modes consistently outperform the BaM baseline across all configurations. When only one queue pair is used, the AGILE async mode provides only marginal speedup over the AGILE sync mode. This phenomenon arises because the number of available SQEs is too small to support all the requests issued in an epoch. As a result, in the prefetch stage, the threads must wait until the AGILE service receives completions from the SSD and recycles SQEs. Consequently, this waiting degrades the asynchronous mode, causing it to exhibit a similar behavior to the synchronous mode in AGILE. As the number of queue pairs increases, more SQEs are available for each epoch. This reduces contention during the prefetch stage and allows the prefetch stage to proceed without stalls. Therefore, the speedup of AGILE async over the synchronous mode becomes more significant.

\begin{figure}[tbh]
\vspace{-10pt}
    \centering
    \includegraphics[width=1\linewidth]{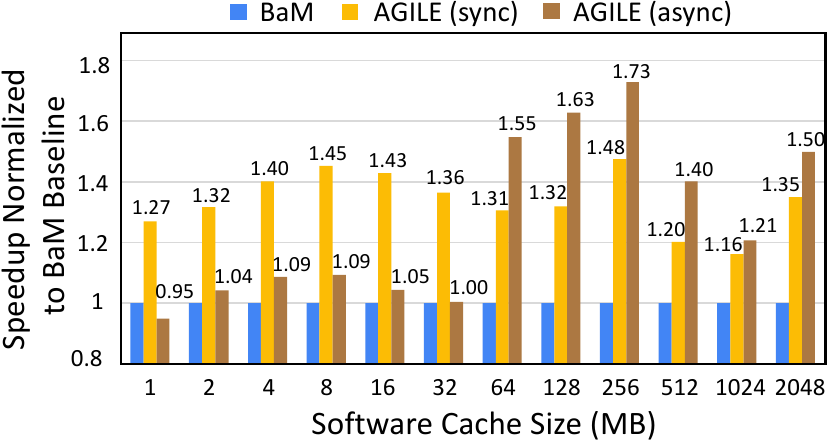}
    \vspace{-20pt}
    \caption{Speedup comparison of AGILE (async and sync modes) and BaM under varying software cache sizes in DLRM inference.}
    \label{fig:dlrm-cache}
    \vspace{-15pt}
\end{figure}

Lastly, we evaluate the impact of software cache size on the DLRM inference using DLRM Config-1 and a batch size of 2048. 
We sweep the software cache size from 1 MB to 2 GB and compare the speedup of AGILE against the BaM baseline. 
Figure~\ref{fig:dlrm-cache} illustrates the changes in the speedup under different software cache sizes. 
The AGILE sync mode consistently outperforms BaM across all cache sizes, achieving a peak speedup of 1.48$\times$ at 256 MB software cache size. 
In contrast, AGILE async mode initially lags behind both the BaM baseline and the AGILE sync mode when the software cache size is small. 
However, the AGILE async mode surpasses the synchronous mode after the software cache reaches a certain threshold, around 64 MB. 
This behavior stems from the software cache contention. 
When the software cache is too small, each epoch may access more data that cannot fit in the software cache size. 
In this case, the prefetch stage in AGILE async will not only wait for available cache lines to make new requests but also evict the previously requested data intended for the next epoch. 
Therefore, when that data is needed in the next epoch, it has already been evicted, and additional requests become necessary during the computation phase. 
The delays in the prefetch stage degrade the asynchronous mode to behave more like the synchronous version, and the extra requests during the computation phase make the performance worse. 
As the software cache size keeps increasing, more cache lines are available to support concurrent prefetching without evictions. 
This allows the prefetch stage to complete soon after the commands are issued. 
Therefore, the data movement time can be hidden by the computation again, exhibiting consistent speedup over the synchronous mode again. 
These results indicate that the asynchronous mode does not always outperform the synchronous one because an improper software cache size will cause stalls and introduce extra NVMe commands. 
Therefore, when applying asynchronous mode in real-world applications, it is essential to estimate both the capacity of the software cache size and the data access demands per epoch to fully leverage the benefits of asynchronous mode.

\subsection{Evaluate AGILE API overhead on graph applications}

The overhead resulting from the implementation is also an important factor that influences overall performance. 
We evaluate the AGILE's API overhead covering both the software cache access and request issuing against BaM on two graph applications: Breadth-First Search (BFS) and sparse matrix vector multiplication (SpMV). The execution time for both BFS and SpMV is dominated by data movement due to their irregular access patterns and low arithmetic intensity~\cite{davis2012spmv, bulucc2011parallel}, making them appropriate benchmarks for assessing API-level overhead. In our experiments, we implement the baseline versions of BFS and SpMV using BaM and AGILE without any application-level optimization, which ensures that the observed performance differences are attributed solely to the underlying software infrastructure, including the API overhead, cache access behavior, and request issuing \& completion polling mechanism.
We use GAP Benchmark Suite~\cite{gapbs} to generate the uniform random graphs and Kronecker graphs to emulate realistic graphs. 
All graph structures and weights (if applicable) are stored in the compressed sparse row (CSR) format. 

To measure the API overhead, we conduct the following three-step experiment:
\begin{enumerate}
    \item We first measure the execution times of the application kernels without using BaM or AGILE, and the graph data is directly stored inside HBM and accessed using the native CUDA API. (Kernel time)
    \item Then, we integrate BaM and AGILE into the application kernels and measure the total runtime, which includes the data transfer time and the overhead from both software cache access and NVMe command issuing. (I/O API time)
    \item Finally, to obtain the overhead in software cache access, we preload all graph data into the software cache before kernel execution, eliminating the NVMe requests during runtime. (Cache API time)
\end{enumerate}

\begin{figure}[tbh]
\vspace{-10pt}
    \centering
    \includegraphics[width=1\linewidth]{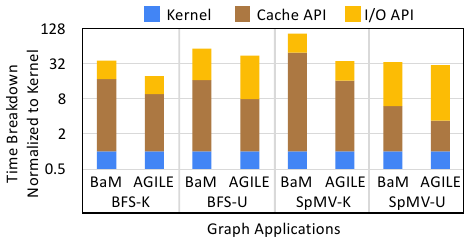}
    \vspace{-20pt}
    \caption{Execution time breakdown of BaM and AGILE across various graph applications.}
    \label{fig:graph-api}
    \vspace{-10pt}
\end{figure}

Figure~\ref{fig:graph-api} illustrates the execution time breakdown of BFS and SpMV using different graph types, where `-K' denotes the Kronecker graphs (K-graph) with skewed degree distribution, and `-U' denotes uniform random graphs (U-graph) with regular structures. The bars are segmented into kernel execution, cache API, and I/O API time. 
All measured execution times are normalized to the kernel runtime. 
Across all graph types, AGILE consistently achieves lower execution time compared with BaM by effectively reducing both the cache API and I/O API overheads. For BFS, AGILE reduces the software cache overhead by 2.27$\times$ on U-graph and 1.93$\times$ on K-graph. and cuts the I/O API overhead by 1.16$\times$ and 1.86$\times$, respectively. For SpMV, AGILE achieves even greater reductions -- 2.11$\times$ and 3.17$\times$ in software cache overhead, and 1.06$\times$ and 2.85$\times$ in I/O overhead on U-graph and K-graph, respectively. These results underscore AGILE's efficiency in handling memory-intensive workloads by minimizing the overhead from the API implementation regardless of graph structure.
\subsection{Evaluate AGILE per thread register usage across CUDA kernels}

\begin{figure}[tbh]
\vspace{-10pt}
    \centering
    \includegraphics[width=0.6\linewidth]{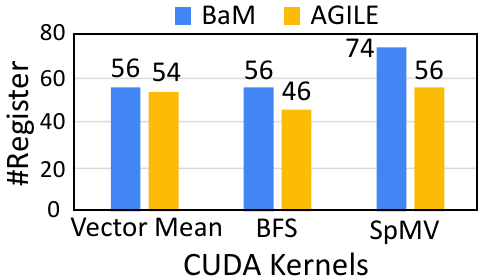}
    \vspace{-10pt}
    \caption{Per-thread register usage comparison between BaM and AGILE across various CUDA kernels.}
    \label{fig:register}
    \vspace{-15pt}
\end{figure}

To further evaluate AGILE's efficiency on GPU resources, we examine its per-thread register usage across different CUDA kernels. Since register usage directly affects warp occupancy and scheduling flexibility, optimizing it is crucial on GPUs. Figure~\ref{fig:register} depicts the number of registers used per thread in different CUDA kernels implemented using BaM or AGILE. We do not impose any constraints to limit the register usage, and both BaM and AGILE use identical kernel implementations for fair comparison.

Compared to BaM, AGILE achieves a reduction in per-thread register by 1.04$\times$, 1.22$\times$, and 1.32$\times$ in Vector Mean, BFS, and SpMV kernels, respectively. This improvement stems from the efficient implementation of AGILE and the offloading of the CQ polling logic to the dedicated AGILE service kernel, which alleviates pressure on application kernels and enables more efficient register utilization. Moreover, the AGILE service kernel is lightweight, which consumes 37 registers per thread and can assist multiple CUDA kernels simultaneously.

\section{Discussion}
While AGILE demonstrates significant performance improvement over existing work and exhibits strong scalability with multiple SSDs, several opportunities remain for extending AGILE to broader and more complex system architectures.

First, extending the software cache hierarchy to incorporate CPU DRAM as an additional tier is a natural and well-motivated enhancement, as demonstrated in prior work~\cite{liu2017hardware, hong2024bandwidth, chang2024gmt}. AGILE is designed with the flexibility to support such an extension. In its current implementation, AGILE includes reserved APIs that enable integration of CPU DRAM as an additional level of the software-managed cache, complementing the existing GPU HBM cache. We will optimize and incorporate this functionality in our open-sourced GitHub repository soon.

Second, AGILE currently targets a single-GPU with multiple SSDs scenario, but AGILE has all the capabilities to be extended to support multiple GPUs with multiple SSDs. To simply share one SSD among GPUs, different I/O queue pairs of the target SSD can work independently and be assigned to different GPUs. It only requires some modifications to the Host APIs, while the AGILE service and interfaces on the CUDA kernel do not need any change. Allowing one GPU to issue peer-to-peer data transfers between another GPU and SSDs or populating data from one GPU directly to another GPU is also doable if the GPU knows the PCIe BARs of the other GPUs. However, it may require further investigation and optimization to handle data transfer and synchronization efficiently without performance degradation. In a multi-GPU system, enabling one GPU to view other GPUs' HBM as a remote cache and leverage NVLink to transfer cached data may also be worth investigating. This additional cache level in HBMs (shared among GPUs) needs further study on the cache coherency among GPUs, which involves dealing with the cache line metadata and analyzing its performance benefits.

Third, extending AGILE to support more heterogeneous systems with accelerators such as FPGAs could provide more performance gains on diverse workloads with various computation and IO characteristics. For example, by leveraging the FPGA's flexibility and advantages in network processing, FpgaNIC~\cite{wang2022fpganic} develops a GPU-oriented SmartNIC on FPGA to accelerate a broad range of distributed applications on distributed GPUs. Besides, FPGA exhibits good energy efficiency as hardware accelerators~\cite{iccad23aim, fpga23charm, dac23automm, fpga24ssr}, and is a good fit for real-time systems, where determinism is critical~\cite{rtss25clare, dong2024eq, glsvlsi25art}. Collaboration between FPGAs and GPUs may offer both high throughput and lower energy consumption while meeting stringent deadline requirements. Such an extension, however, introduces new challenges in coordinating multiple devices and requires more sophisticated system designs. We leave this extension to future versions of AGILE.

Fourth, AGILE may also enable new research in compiler-level optimizations. For applications involving multiple data communications and computations within a single kernel, the programmers need to explore the overlap opportunity manually. While currently AGILE functions as an asynchronous I/O library, it can be extended with compiler support to automatically analyze dependencies and perform code transformations. Existing research works have explored similar optimizations. For example, the compiler identifies the data dependency and reorders instructions for better overlapping~\cite{crago2024wasp, ariesfpga25}. AGILE serves as a foundational first step toward that goal of developing a compiler that enables static dependency analysis to automatically explore overlapping opportunities.

Fifth, supporting AGILE in virtualized environments, such as virtual machines or Docker containers, is important for improving portability, scalability, and ease of deployment in shared computing infrastructures. However, this requires further development and investigation into the associated performance implications, particularly with respect to I/O virtualization, device passthrough, and potential overhead introduced by the virtualization layer. 

\vspace{-5pt}
\section{Conclusion}
In this paper, we propose AGILE, a lightweight and efficient asynchronous library for GPU-SSD integration. AGILE is the first work that enables GPU threads to issue NVMe commands asynchronously and allows users to customize software cache policy. AGILE enables overlapping at the thread level and achieves up to 1.88$\times$ reduction in execution time by hiding data transfer with computation. AGILE exhibits up to 1.75$\times$ improvement on DLRMs and shows 3.12$\times$ and 2.85$\times$ API overhead reduction in software cache and NVMe IO requests compared with the state-of-the-art GPU-centric work, BaM~\cite{qureshi2023gpu}. AGILE is also lightweight and consumes up to 1.32$\times$ fewer registers in various CUDA kernels.

\smallskip
{
{\noindent\textbf{ACKNOWLEDGEMENTS --}} This work is supported in part by Brown University New Faculty Start-up Grant, and NSF awards \#2213701, \#2217003, \#2328972, \#2511445, \#2536952.  
\bibliographystyle{ACM-Reference-Format}
\bibliography{reference}

\appendix

\end{document}